\documentclass[aps,showpacs,twocolumn]{revtex4}

\usepackage{amssymb}
\usepackage{graphicx}
\usepackage{dcolumn}
\usepackage{amsmath}
\newcommand{\kk}{\mathbf{k}}

\begin{document}

\title{Electron states and magnetic phase diagrams of strongly correlated systems}

\author{V. Yu. Irkhin$^{1}$ and P. A. Igoshev$^{1,2}$}
\affiliation
{$^{1}$M. N. Mikheev Institute of Metal Physics, 620108 Ekaterinburg, Russia
}
\affiliation
{$^{2}$Ural Federal University, 620002 Ekaterinburg, Russia}

\begin{abstract}
Various auxiliary-particle approaches to treat electron correlations in many-electron models are analyzed. Applications to copper-oxide layered systems are discussed. The ground-state magnetic phase diagrams are considered within the Hubbard and $s$-$d$ exchange (Kondo) models for square and simple cubic lattices vs. band filling and interaction parameter.
A generalized Hartree-Fock approximation is employed to treat commensurate ferro-, antiferromagnetic, and incommensurate (spiral) magnetic phases, and also magnetic phase separation. The correlations are taken into account within the Hubbard model by using the slave-boson approach. The main advantage of this approach is correct estimating the contribution of  doubly occupied states number and therefore  the paramagnetic phase energy.


\end{abstract}


\maketitle

Magnetic properties of strongly correlated transition-metal compounds and their relation to doping, lattice geometry, band structure and interaction parameters are still being extensively investigated. In particular, the details of magnetic order in the ground state remain to be examined both theoretically and experimentally. During recent
decades, the two-dimensional (2D) case closely related to the problem of high-temperature superconductivity in cuprates and iron arsenides has been intensively investigated theoretically.

The ground state of strongly correlated systems is characterized by a competition  of ferromagnetic (FM) and antiferromagnetic (AFM) ordering which results in occurrence of  spiral magnetic ordering    \cite{Igoshev:2010} or the magnetic phase separation \cite{Visscher:1973, Igoshev:2010, Igoshev:2015}. The consideration of these problems is performed within a number of many-electron models. In the present work we discuss theoretical approaches to treat these model within auxiliary-particle representations (Sect.1) and present some results of numerical calculations (Sect.2).

\section{Theoretical models and slave particle representations}

To describe the properties of such systems one uses many-electron models like the Hubbard, $s$-$d$ exchange (Kondo) and Anderson lattice models. These  are widely applied, {\it e.~g.}, for high-$T_c$ cuprates and rare earth compounds. 
There exist some relations (mappings) between these models in various parameter regions.

The Hamiltonian of the Hubbard model reads
\begin{equation}
      \label{eq:original_H}
      \mathcal{H}_{\rm H} = \sum_{ij\sigma} t_{ij}c^\dag_{i\sigma}c^{}_{j\sigma}+U\sum_i n_{i\uparrow}n_{i\downarrow},
\end{equation}
where $c^{\dagger}_{i\sigma}$ are electron creation  operators. 
In the limit of large Hubbard parameter $U$ and band filling $n<1$ (hole doping) this is reduced to the $t-J$ model
\begin{equation}
\mathcal{H}_{t_J}=-\sum_{ij\sigma }t_{ij}X_i(0\sigma )X_j(\sigma 0)
+ \sum_{ij}J_{ij}\mathbf{S}_i\mathbf{S}_j.
 \label{eq:I.7}
\end{equation}
where $X_i(\Gamma ,\Gamma ^{\prime })=|\Gamma_i \rangle \langle \Gamma
^{\prime }_i|$  are the Hubbard X-operators acting on the $i$ site local subspace \cite{654}, $J_{ij} = 2t^2_{ij}/U$.


To proceed with analytical and numerical calculations, it is convenient to use auxiliary  (``slave'') boson and fermion representations.
In connection with the theory of high-temperature
superconductors, Anderson \cite{633a} put forward
the idea of the separation of the spin and charge
degrees of freedom of electron ($\sigma = \pm1$):
\begin{equation}
c_{i\sigma }^{\dagger
}=X_i(\sigma ,0)+\sigma X_i(2,-\sigma )\rightarrow f_{i\sigma }^{\dagger
}e_i+\sigma d_i^{\dagger }f_{i-\sigma }.
 \label{eq:6.131}
\end{equation}
Here, $f_{i\sigma }^{\dagger }$ are the creation operators for neutral fermions (spinons), and $e_i^{\dagger }$, $d_i^{\dagger}$ are the creation operators
for charged spinless bosons (holons and doublons). For large $U$, we have to retain only holon or doublon degrees of freedom for hole or electron doping, respectively.

In fact, the choice of the Fermi statistics for spinons and Bose one for holons is not unique and can be varied depending on the physical picture ({\it e.g.},~presence or absence of magnetic ordering, see also \cite{Kane}).
Thus,  the spin operators in Eq.~(\ref{eq:I.7}) are presented as the bilinear form 
\begin{equation}
\mathbf{S}_i=\frac 12\sum_{\sigma \sigma ^{\prime }}a_{i\sigma
}^{\dagger }\mbox{\boldmath$\sigma $}_{\sigma \sigma ^{\prime
}}a_{i\sigma ^{\prime }}, \,
 \label{eq:O.1}
\end{equation}
in terms of Schwinger bosons ($a_{i\sigma }^{\dagger } = b_{i\sigma }^{\dagger }$) or fermionic spinons ($a_{i\sigma }^{\dagger } = f_{i\sigma}^{\dagger }$), $\mbox{\boldmath$\sigma $}$ being Pauli matrices.

A more complicated representation proposed by Kotliar and Ruckenstein~\cite{Kotliar86}
introduces the slave boson operators $e_{i},\,p_{i\sigma },\,d_{i}$, so that
\begin{equation}
c_{i\sigma }^{\dag }\rightarrow f_{i\sigma }^{\dag }z_{i\sigma }^{\dag }
\end{equation}
where $f_{i\sigma },f_{i\sigma }^{\dag }$ are the slave Fermi operators, and
\begin{multline}
z_{i\sigma }^{\dag}=(1-e_{i}^{\dag }e_{i}-p_{i-\sigma}^{\dag }p_{i-\sigma})^{-1/2}(p^{\dag }_{i\sigma }e^{}_{i} + d^{\dag}_{i}p^{}_{i-\sigma})
\\
\times (1-d_{i}^{\dag }d_{i}-p_{i\sigma }^{\dag }p_{i\sigma
})^{-1/2}.
\label{zzz}
\end{multline}%

There exists also a rotationally invariant version \cite{Fresard:1992a}
\begin{equation}
c_{i\sigma }^{\dag }=\sum_{\sigma ^{\prime }}f_{i\sigma }^{\dag }z_{i\sigma
\sigma ^{\prime }}^{\dag },~\hat{z}_{i}=e_{i}^{\dag}\hat{L}_{i}M_{i}\hat{R}_{i}%
\hat{p}_{i}+\widehat{\tilde{p}}_{i}^{\dag}\hat{L}_{i}M_{i}\hat{R}_{\hat{i}%
}d_{i}
\label{zz}
\end{equation}%
where the factors $L,M,R$  are similar to those in (\ref{zzz}), the scalar and vector slave
boson fields $p_{i0}$ and $\mathbf{p}_{i}$ are introduced as $\hat{p}_{i}=\frac{1%
}{2}(p_{i0}\sigma _{0}+\mathbf{p}_{i}\mbox{\boldmath$\sigma $})$  and $%
\widehat{\tilde{p}}_{i}$ is the time reverse of operator $\hat{p}_{i}$.
  This version is suitable for magnetically ordered phases to  take into account spin fluctuation corrections. In particular, it can describe in a simple way non-quasiparticle states owing electron-magnon scattering which were earlier treated in the many-electron representation of Hubbard's operators (cf. Ref.\cite{338}).


We mention also the rotor representation \cite{Florens}
\begin{equation}
c_{i\sigma }^{\dag }=f_{i\sigma }^{\dag }\exp (i\theta _{i})
\end{equation}
where the original Hubbard interaction is replaced in by a simple kinetic
term for the phase field, $(U/2)\hat{L}_{i}^{2}$, with the angular momentum $%
\hat{L}=-i\partial /\partial \theta $. This representation is suitable to
describe the metal-insulator transition in the paramagnetic phase.

In the case of the Anderson lattice model transport and magnetic properties are separated between different systems, $s$ and $d$ correspondingly: 
\begin{multline}\label{eq:hamiltonian of Anderson}
\mathcal{H}_{\rm A} = \sum_{ij\sigma} t_{ij}c^{\dagger}_{i\sigma}c_{j\sigma} + \epsilon_d\sum_{i\sigma}d^{\dagger}_{i\sigma}d^{}_{i\sigma} \\
 + V\sum_{i\sigma}(c^{\dagger}_{i\sigma}d^{}_{i\sigma} + d^{\dagger}_{i\sigma}c^{}_{i\sigma}) +
U\sum_i d^{\dagger}_{i\uparrow}d^{}_{i\uparrow}d^{\dagger}_{i\downarrow}d^{}_{i\downarrow},
\end{multline}
$\epsilon_d$ is the energy of localized (`$d$-electron') state,
$V$ is on-site $s$-$d$ hybridization providing the coupling between these subsystems.

Provided that the $d$-level is well below the Fermi energy and Coulomb interaction is sufficiently large ($|V|\ll \epsilon_{\rm F}-\epsilon_d$, $U$), this model can be reduced by the Schrieffer-Wolff transformation  to the $s$-$d$ exchange model
\begin{equation}\label{eq:sd_model_def}
	\mathcal{H}_{s-d} = \sum_{\kk\sigma} t_{\kk}c^\dag_{\kk\sigma}c^{}_{\kk\sigma} - I\sum_{i \sigma\sigma'} (\mathbf{S}_i\cdot\mbox{\boldmath$\sigma $}_{\sigma\sigma'}) c^\dag_{i\sigma}c^{}_{i\sigma'},
\end{equation}
with spin $S=1/2$ and the exchange parameter
\begin{equation}
	I = V^2[1/(\epsilon_d-\epsilon_{\rm F}) - 1/(U+\epsilon_d-\epsilon_{\rm F})],
\end{equation}
where $\epsilon_{\rm F}$ is the Fermi level.

Remember also the three-band model of cuprates
\begin{eqnarray}
\mathcal{H}&=&\sum_{\mathbf{k}a\sigma }\left[ \varepsilon
p_{\mathbf{k}a\sigma }^{\dagger }p_{\mathbf{k}a\sigma }+\Delta
d_{\mathbf{k}\sigma }^{\dagger }d_{\mathbf{k}\sigma
}+V_{\mathbf{k}}(p_{\mathbf{k}\sigma }^{\dagger
}d_{\mathbf{k}\sigma }+h.c.)\right]
\nonumber\\
&+&U\sum_id_{i\uparrow }^{\dagger }d_{i\uparrow }d_{i\downarrow
}^{\dagger }d_{i\downarrow },
 \label{eq:6.108}
\end{eqnarray}
where $\varepsilon $ and $\Delta $ are positions of  $p$- and $d$-levels
for O- and  Cu-ions, $V_{\mathbf{k}}=2V_{pd}\left( \sin ^2k_x+\sin ^2k_y\right) ^{1/2}$ are the matrix elements of  $p$---$d$ hybridization (cf. \cite{Raimondi}).
In the large-$U$ limit we can use the slave boson representation $d^\dag_{i\sigma} \rightarrow X{_i(\sigma 0) }=f^{\dag}_{i\sigma}e_i$.

For $|V_{pd}|\ll \varepsilon -\Delta $ (large charge-transfer gap)
the Hamiltonian~(\ref{eq:6.108}) is again reduced by a canonical transformation to the $t-J$ model with $t_{\text{eff}}=V_{pd}^2/(\varepsilon -\Delta )$.
It is interesting that the $t-J$ model obtained from the one-band Hubbard model is also formally reduced to a similar structure in the representation (\ref{zz}) with auxiliary rather than physical particles $p_i$. Thus the Hubbard model and the model (\ref{eq:6.108}) can be considered in a parallel way \cite{Raimondi}.

To describe doped cuprates, also a representation of the Fermi dopons $d^\dag_{i\sigma
}$ was proposed \cite{Ribeiro,Scr}
\begin{equation}
X{_i(0, -\sigma) }=-\frac \sigma{2\sqrt{2}}\sum_{\sigma ^{\prime }}d_{i\sigma
^{\prime }}^{\dagger }(1-n_{i,-\sigma ^{\prime }})
[\delta _{\sigma \sigma ^{\prime }}-2(\mathbf{S}_i\mbox{\boldmath$\sigma $}_{\sigma ^{\prime }\sigma} )].
 \label{eq:I.8}
\end{equation}
On substituting (\ref{eq:I.8}) into the $t-J$ Hamiltonian (2) we obtain the terms  which are linear in spin operators. These can provide hybridization between electrons (dopons) and Fermi spinons to describe nodal-antinodal dichotomy  and formation of large Fermi surface in cuprates with  the increase of doping \cite{Scr}. Thus the initial one-band model takes the form of an effective two-band model. Note that a similar structure is obtained in the spin-rotation representation (\ref{zz}).

On the other hand, in the antiferromagnetic case it is
convenient to  use the Schwinger rather than the spinon representation
(see (\ref{eq:O.1})). Such an approach was developed in \cite{Punk}  to describe the formation of spin liquid state in terms of frustrations in localized-spin subsystems.
The supersymmetry approach  \cite{Lavagna} mixes the fermionic and the bosonic representation of the spin following the standard rules of superalgebra.

\section{Results of numerical calculations}

After the local rotation in the
spin space matching the local magnetization vectors at different sites, say along $z$ axis,
(which is necessary for the consideration of magnetic spirals) by the angle $%
\mathbf{QR}_{i}$ (where $\mathbf{Q}$ is the wave vector of the spiral) the Hubbard Hamiltonian in the slave boson representation (\ref{zzz}) takes the form
\begin{equation}
\mathcal{H}_{\mathrm{eff}}=\sum_{ij\sigma \sigma ^{\prime }}t_{ij}^{\sigma
\sigma ^{\prime }}f_{i\sigma }^{\dag }f_{j\sigma ^{\prime }}z_{i\sigma
}^{\dag }z_{j\sigma ^{\prime }}+U\sum_{i}d_{i}^{\dag }d_{i},
\end{equation}%
with  $t_{ij}^{\sigma \sigma ^{\prime }}=\exp [\mathrm{i}\mathbf{Q}(\mathbf{R}_{i}-%
\mathbf{R}_{j})\sigma ^{x}]_{\sigma \sigma ^{\prime }}t_{ij}$.
This form enables us to construct the mean-field approximation where the boson averages do not depend on lattice site.

\begin{figure}[h!]
\includegraphics[width=0.5\textwidth]{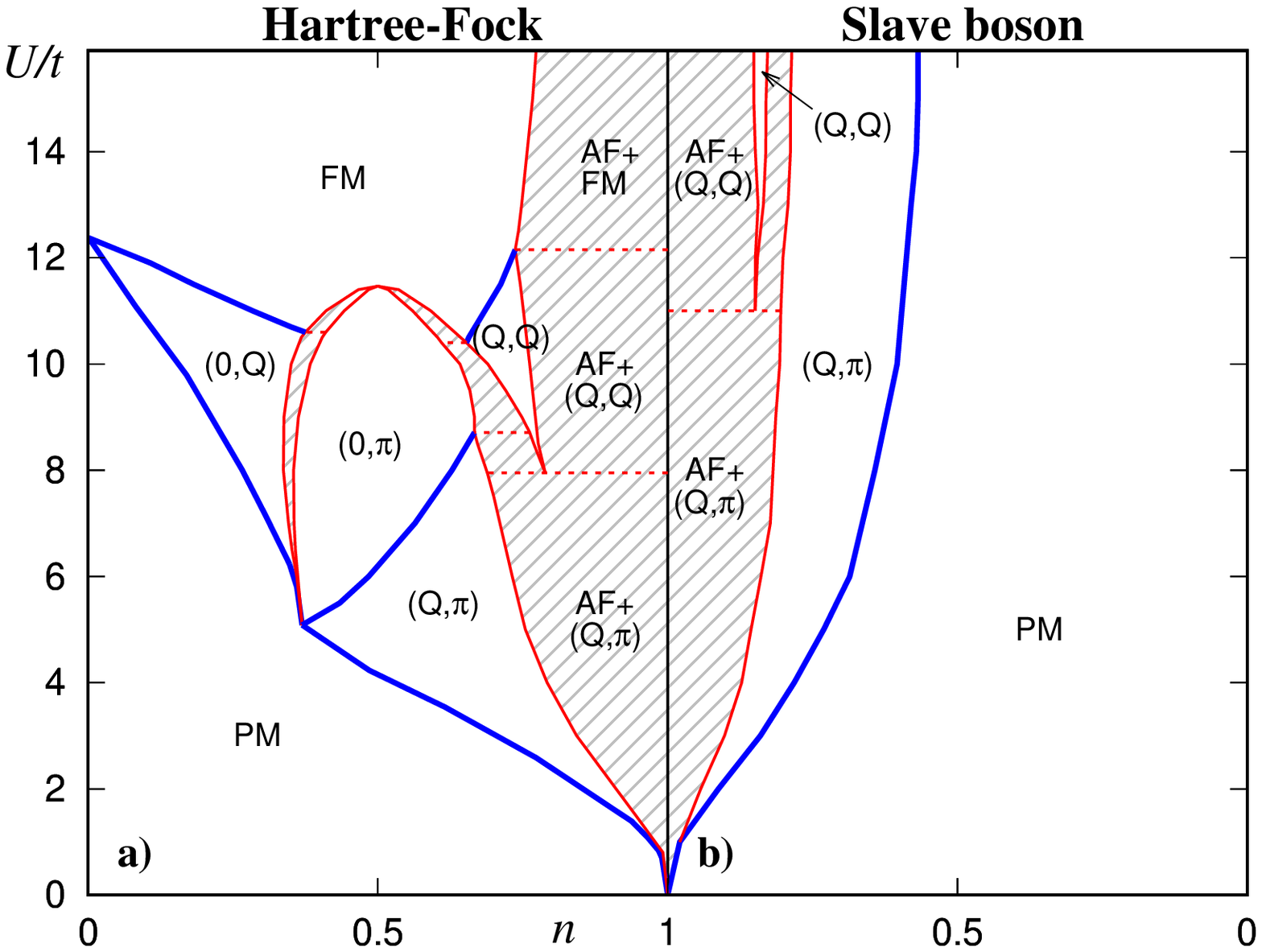}
\includegraphics[width=0.49\textwidth]{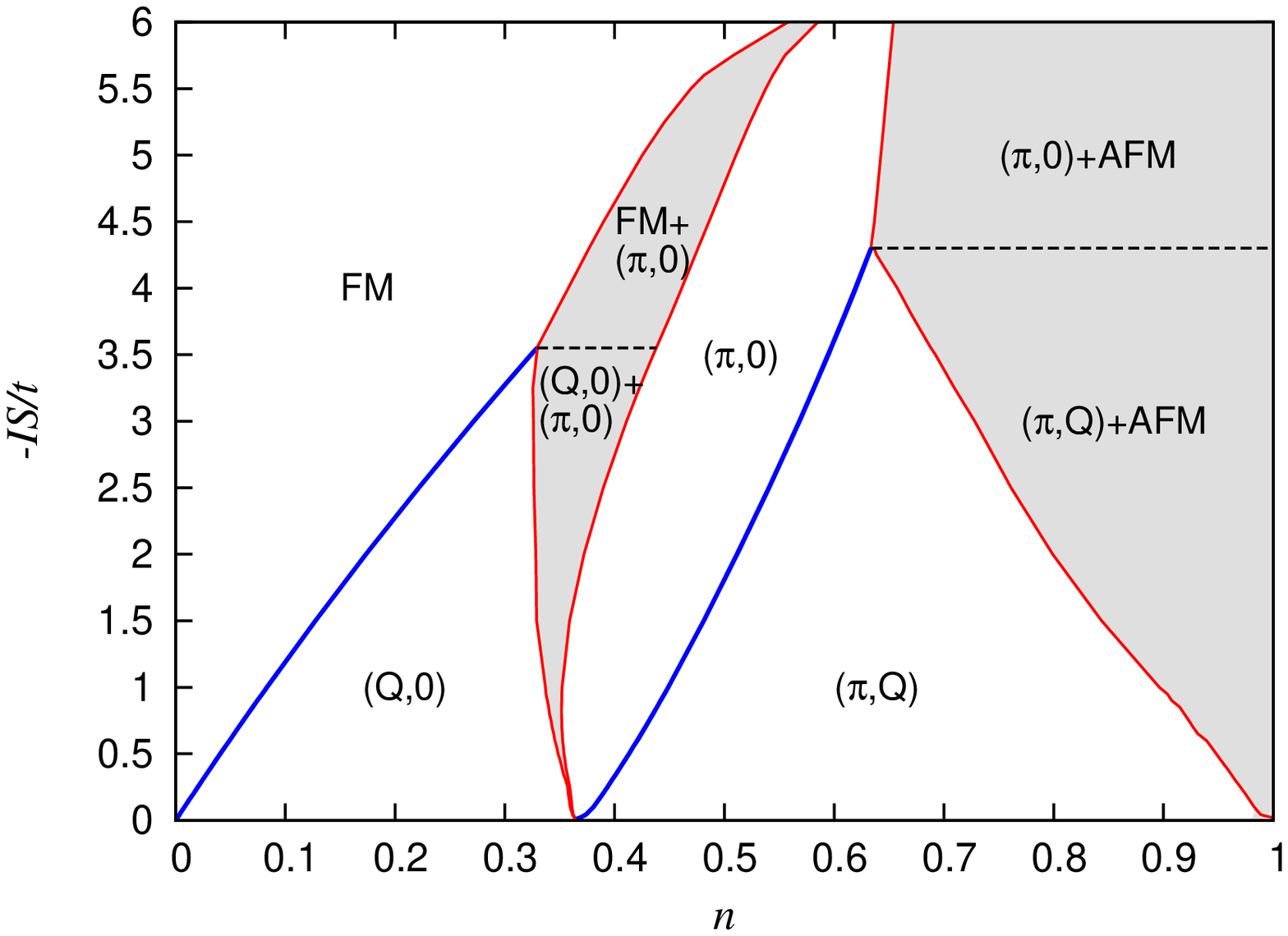}
\caption{
	(Color online)
	Ground state magnetic phase diagram of the Hubbard model (upper panel, within (a) HFA and (b) SBA) and $s$-$d$ model (lower panel) for the square lattice 
	at $n<1$.
	The spiral phases are denoted according to the form of their wave vector.
	Filling shows the phase separation regions, dashed (red) lines being the boundaries  between different phase pairs. 	Bold (blue) lines denote the second-order phase transitions.
	Solid (red) lines correspond to the boundaries between  a homogeneous phase and phase separation region,
	$\mathbf{Q}_{\rm AFM} = (\pi,\pi)$.
}
\label{fig:sq}
\end{figure}
\begin{figure}[h!]
\includegraphics[width=0.49\textwidth]{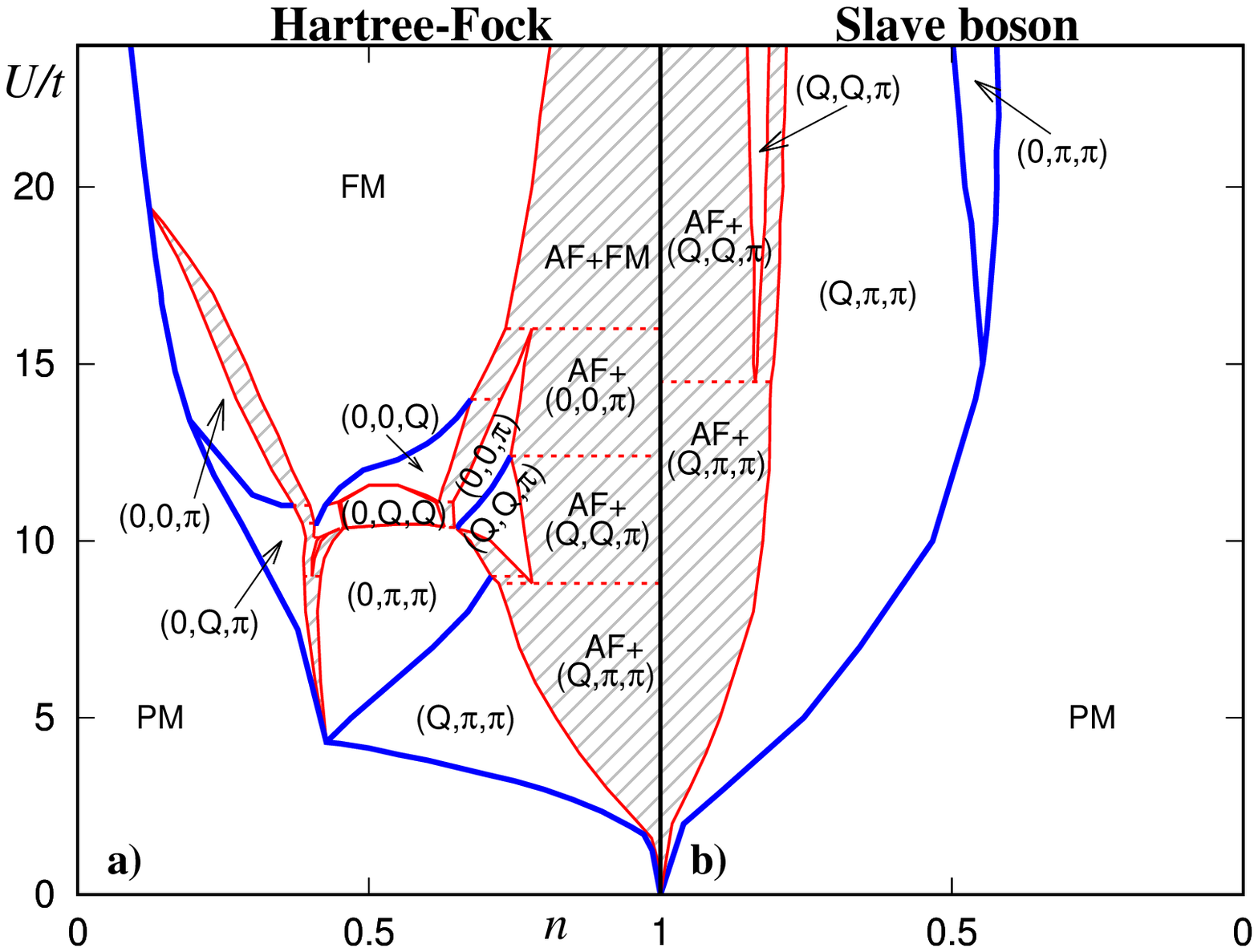}
\includegraphics[width=0.49\textwidth]{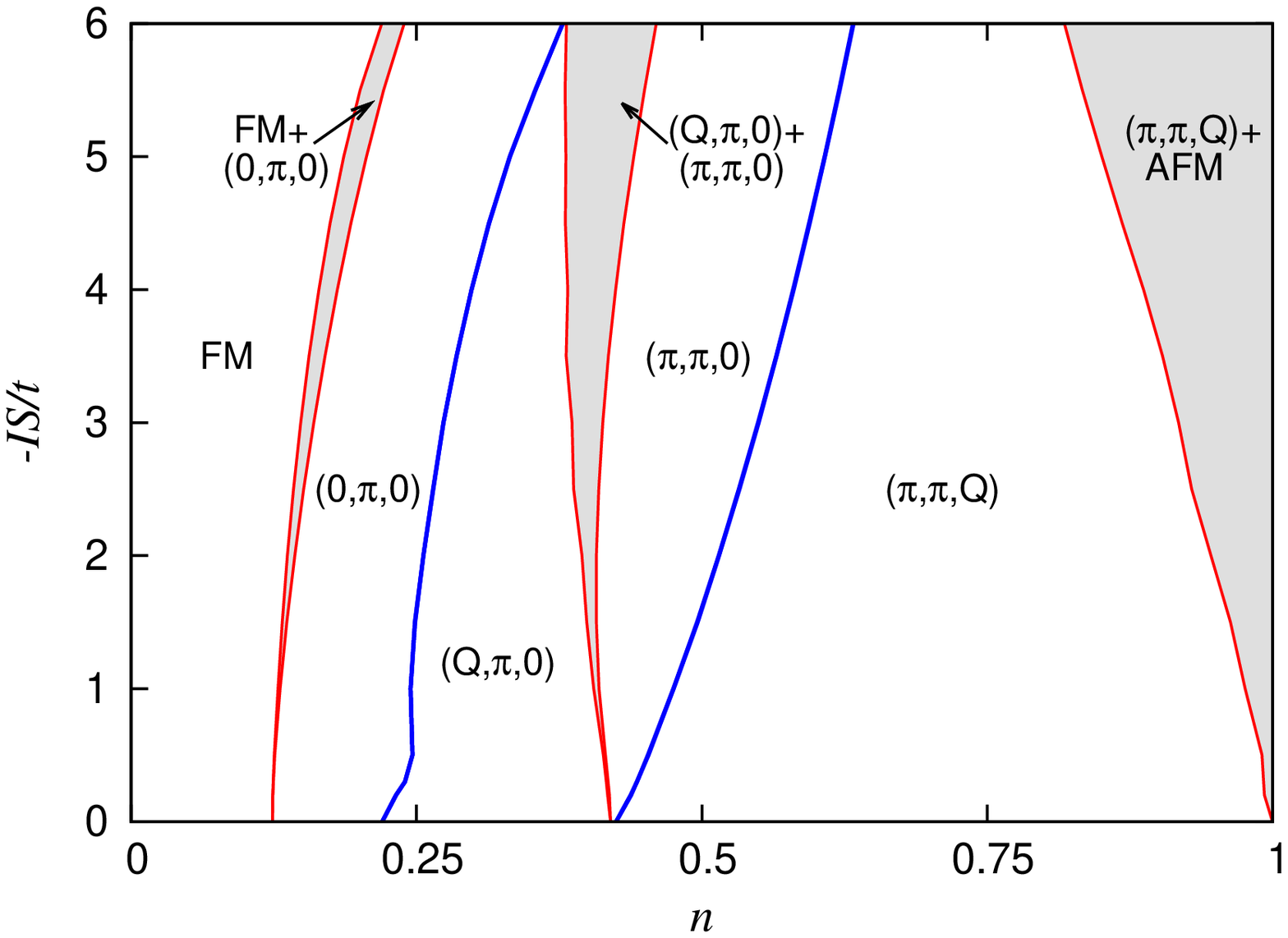}
\caption{
	(Color online)
	The phase diagrams for the simple cubic  lattice. Notations are the same as in Fig. \ref{fig:sq}, $\mathbf{Q}_{\rm AFM} = (\pi,\pi, \pi)$. 
}
\label{fig:sc}
\end{figure}

The calculations were performed for the half-filled Hubbard model to describe the metal-insulator transition and for the magnetic phase diagram for arbitrary filling \cite{Timirgazin,Timirgazin1,Igoshev:2015}.
Here we present the results for the square~(Fig.~\ref{fig:sq}) and simple cubic~(Fig.~\ref{fig:sc}) lattices in the Hubbard model within the Hartree--Fock approximations  (HFA) and slave boson approach (SBA)  in the mean-field version and within HFA for the $s$-$d$ exchange model \cite{Igoshev:2015,Antipin}.

One can see that  HFA yields the variety of spiral magnetic phases, as well as FM and AFM ones at sufficiently large $U$ generally and at any $U$ in the vicinity of half-filling.
The account of correlations (SBA) 
leads to a noticeable suppression of magnetically ordered states in comparison
with HFA: the corresponding density intervals in the phase diagram decrease strongly, and the variety of the spiral states disappears.
Besides that, in SBA there occurs a wide region of paramagnetic (PM) state which is a manifestation of correct treatment of the energy of doubles.

It should be stressed that only SBA provides a correct description of large $U$ case at finite density of current carriers, whereas the ground state energy in HFA is divergent at $U \rightarrow \infty$ due to overestimation of Coulomb interaction energy, as well as in random phase approximation (RPA) \cite{353}.

Within the mean-field approximation, the Hubbard model is equivalent to the $s$-$d$ model with the replacement $IS = Um/2$. However, the phase separation condition and description of the PM phase are different in these models. Thus  the phases are strongly redistributed.

Due to existence of localized moments,  ferromagnetic ordering in the $s$-$d$ model is favorable
already at small $|I|$, whereas within the Hubbard model it occurs at sufficiently large $U$ only.
The increase of $|I|$ results in a growith of FM region.
At small $|I|$ the wave vector of magnetic phase is specified by the position of the maximum of the Lindhardt function
\begin{equation}\label{eq:sd_model_diagram}
	\chi^0_{\mathbf{Q}} = \frac1{N}\sum_{\kk}\frac{f_{\mathbf{k}+\mathbf{Q}}-f_{\mathbf{k}}}{t_{\mathbf{k}}-t_{\mathbf{k}+\mathbf{Q}}}
\end{equation}
calculated in the PM phase ($f_{\mathbf{k}}$ is the Fermi function).

An important difference of the square and simple cubic lattice (2D and 3D) cases is the form of FM region at small $n$. In the former case its width vanishes at $I\rightarrow 0$, but in the latter case the width is finite and sufficiently large, the transition to the spiral $(0,\pi,0)$ phase being of the first order. 
For the square lattice, the spiral phases are fully suppressed by FM and AFM regions at $|IS|\gtrsim6t$.
For the simple cubic lattice, the spiral phases turn out to be more stable.
For small number of carriers in the AFM matrix ($1-n\ll 1$),  the phase separation between AFM and spiral  ($(Q,\pi)$ for square lattice and $(Q,\pi, \pi)$ for simple cubic lattice) phases is present at small interaction parameter.


The research was carried out within the state assignment of FASO of Russia (theme ``Quantum'' No. AAAA-A18-118020190095-4). This work was supported in part by Ural Division of RAS (project no. 18-2-2-11) and by the Russian Foundation for Basic Research (project no.
16-02-00995).

\end{document}